\documentclass[webpdf,contemporary,large,namedate]{oup-authoring-template}%

\usepackage{lmodern}
\usepackage{algorithm}
\usepackage[noEnd=true,commentColor=darkgray]{algpseudocodex}
\tikzset{algpxIndentLine/.style={draw=darkgray}}
\usepackage{setspace}

\begin{document}
\journaltitle{TBD}
\DOI{TBD}
\copyrightyear{2026}
\pubyear{2026}
\access{Advance Access Publication Date: Day Month Year}
\appnotes{Preprint}
\firstpage{1}

\title[Fast read alignment with minibwa]{Fast genomic read alignment with minibwa}
\author[1,2,3,$\ast$]{Heng Li\ORCID{0000-0003-4874-2874}}
\author[4]{Nils Homer\ORCID{0009-0007-7860-1155}}
\address[1]{Department of Data Science, Dana-Farber Cancer Institute, 450 Brookline Ave, Boston, MA 02215, USA}
\address[2]{Department of Biomedical Informatics, Harvard Medical School, 10 Shattuck St, Boston, MA 02215, USA}
\address[3]{Broad Institute of MIT and Harvard, 415 Main St, Cambridge, MA 02142, USA}
\address[4]{Fulcrum Genomics LLC, 240 Elm Street, Somerville, MA, 02144, USA}
\corresp[$\ast$]{Corresponding author. \href{mailto:hli@ds.dfci.harvard.edu}{hli@ds.dfci.harvard.edu}}


\abstract{
\sffamily\footnotesize
\textbf{Motivation:}
BWA-MEM remains a popular short-read mapper especially for the purpose of variant calling.
Several groups have accelerated this algorithm as it has been the performance bottleneck of many current workflows.
However, constrained by the original design,
these drop-in replacements could only achieve limited speedup.
Breaking changes to BWA-MEM are required for further improvement.\vspace{0.5em}\\
\textbf{Results:}
We developed minibwa for aligning short and accurate long reads against a reference genome.
It combines BWA-MEM variable-length seeding with minimap2 chaining and base alignment.
It speeds up BWA-MEM2 further with additional prefetch for seeding,
new heuristics to skip unnecessary mate rescue and reduced effort in highly repetitive regions
where reads would anyway be wrongly mapped due to structural changes.
Minibwa is about four times as fast as BWA-MEM and over twice as fast as BWA-MEM2 at comparable accuracy.
It also natively supports directional bisulfite sequencing data to high mapping accuracy.
\vspace{0.5em}\\
\textbf{Availability and implementation:}
\url{https://github.com/lh3/minibwa}
}

\maketitle

\section{Introduction}

Read alignment is indispensable to most reference-based sequence analyses such as variant calling and methylation profiling.
Although over 100 read mappers have been published~\citep{Alser:2021aa,DBLP:journals/csur/ProusalisGKNKAPSG26},
only several are maintained and remain widely used.
The two most notable short-read mappers for DNA sequencing data are Bowtie2~\citep{Langmead:2012fk} and BWA-MEM~\citep{Li:2013aa}, both based on FM-index~\citep{DBLP:conf/focs/FerraginaM00}.
While Bowtie2 is equally popular for general purposes, BWA-MEM is more often used for variant calling.

BWA-MEM is the primary performance and cost bottleneck in variant calling pipelines.
Given that human Whole-Genome Sequencing (WGS) accounts for the vast majority of sequencing data volume, this inefficiency is particularly pronounced.
There have been several efforts to accelerate the original BWA-MEM algorithm while maintaining identical output.
BWA-MEM2~\citep{DBLP:conf/ipps/VasimuddinMLA19} revamps the memory layout of FM-index
and accelerates extension alignment with Single Instruction Multiple Data (SIMD) instructions.
It is 50--100\% faster than BWA-MEM.
BWA-MEME~\citep{Jung:2022aa} further speeds up BWA-MEM2 seeding with learned indices.
BWA-MEM3 is a new fork of BWA-MEM2 with speedup from latency hiding and additional fast paths to skip full dynamic programming.
Sentieon BWA is a proprietary drop-in replacement of BWA-MEM.
Parabricks~\citep{OConnell:2023aa,Zhu2025.07.23.666378} ports BWA-MEM to GPU.
It does not promise identical output but still produces alignment qualitatively equivalent to BWA-MEM.

However, designed more than a decade ago, BWA-MEM has several limitations that prevent further optimization without changing the output.
First, BWA-MEM uses a complex algorithm to seed alignment~\citep{Li:2012fk} which is slower than a recent algorithm~\citep{DBLP:conf/dlt/Gagie24,Li:2024ac}.
Second, BWA-MEM chaining is heuristic and inferior to the minimap2 chaining algorithm~\citep{Li:2018ab}.
It does not work well for reads longer than 500bp, which will become more common with Roche's SBX sequencing technology~\citep{Kokoris2025.02.19.639056} ---
the boundary between short and long reads is blurring.
Third, BWA-MEM uses a non-SIMD-based extension algorithm with intricate boundary conditions.
This complicates SIMD-based equivalence in BWA-MEM2 and increases maintenance burden.
Fourth, BWA-MEM tries too hard in highly repetitive regions.
Due to rapid evolution in centromeres~\citep{Gao2025.12.09.693231}, 
exact base alignment in centromeres is challenging even with complete assemblies~\citep{Bzikadze:2022aa}.
Exhaustive alignment in centromeres is a waste of effort
as we cannot accurately place centromeric short reads anyway due to large structural differences.
There is still room for improvement if breaking changes can be introduced.

BWA-MEM does not natively support bisulfite sequencing (BS-seq) reads.
BWA-Meth~\citep{Pedersen:2014aa} is a wrapper around stock BWA-MEM that uses the so-called 3-base strategy to align directional BS-seq data.
Without touching the BWA-MEM source code, BWA-Meth wastes compute on unsuccessful extension alignment and breaks the pairing logic.
BISCUIT~\citep{Zhou:2024ac} addresses these issues by customizing the BWA-MEM source code.
As a tradeoff, though, the deep code integration makes it challenging to adopt the faster BWA-MEM2 implementation.
Combining an efficient mapping algorithm with the BISCUIT strategy would yield an even more capable BS-seq mapper.

In this article, we will describe minibwa, the next iteration of BWA-MEM.
Minibwa aligns standard WGS short reads and Hi-C reads like BWA-MEM
and additionally maps accurate long reads like minimap2 and it natively supports BS-seq data.

\section{Methods}

In a nutshell, minibwa is a hybrid of BWA-MEM~\citep{Li:2013aa} and minimap2~\citep{Li:2018ab},
combining BWA-MEM variable-length seeding with minimap2 chaining and SIMD-based alignment.
It additionally adapts the ropebwt3~\citep{Li:2024ac} algorithm for finding supermaximal exact matches (SMEMs).
Table~\ref{tab:relation} summarizes the relationship between minibwa and other algorithms.
BWA-MEM2 also batches suffix array (SA) query and implements SIMD-based alignment.
Further performance gains in minibwa primarily come from the new SMEM algorithm,
more frequent use of ungapped alignment and reduced effort in highly repetitive regions such as centromeres.

\begin{table}[t]
\caption{Relationship with BWA-MEM, ropebwt3 and minimap2\label{tab:relation}}
\begin{tabular*}{\columnwidth}{@{\extracolsep\fill}lll@{\extracolsep\fill}}
\toprule
Component & Source & Modification \\
\midrule
FM-index  & BWA-MEM  & Near identical \\
SMEM      & ropebwt3 & Reimplemented and batched \\
SA query  & ropebwt3 & Batched differently \\
Seeding   & BWA-MEM  & Different algorithm \\
Chaining  & minimap2 & Adapted for variable-length seeds \\
Alignment & minimap2 & More frequent ungapped fast path \\
Pairing   & BWA-MEM  & Ungapped fast path \\
\botrule
\end{tabular*}
\end{table}

For the completeness of the method section, we will describe the basic theory on FM-index in brief.
Additional theoretical background can be found in \citet{Li:2024ac}.
Note that the original BWA paper~\citep{Li:2009uq} used closed intervals but here we adopt half-closed-half-open intervals for consistency with our recent work.

\subsection{Definitions}

$\Sigma\triangleq\{{\tt A},{\tt C},{\tt G},{\tt T}\}$ is the DNA alphabet
with lexicographical order ${\tt A}<{\tt C}<{\tt G}<{\tt T}$.
$P\in\Sigma^*$ is a \emph{string} of length $|P|$.
$P[i]$, $0\le i<|P|$, is the $i$-the symbol and
$P[i,j)$ is the $(j-i)$-long substring starting at position $i$.
Operator $\circ$ concatenates two strings.
It may be omitted if concatenation is apparent in the context.
For $a\in\Sigma$, $\overline{a}$ is the Watson-Crick complement of $a$.
$\overline{P}$ is the reverse complement of string $P$.

$R\in\Sigma^*$ is the reference genome with all contigs concatenated and ambiguous bases converted to random bases.
$T\triangleq R\circ\overline{R}\circ\$$ further concatenates $R$ and its reverse complement
with a trailing \emph{sentinel} $\$$ that is smaller than all symbols in $\Sigma$.
For convenience, let $n=|T|$ and $T[-1]=T[n-1]=\$$.
The \emph{suffix array} (\emph{SA}) of $T$ is an integer array $S$ where
$S(i)$ is the position of the $i$-th smallest suffix of $T$.
$B$ is the \emph{Burrows-Wheeler Transform} (\emph{BWT}) of $T$.
By definition, $B[i]=T[S(i)-1]$.

\begin{algorithm}[b]
	\caption{Backward and forward extension}\label{algo:backfor}
	\begin{algorithmic}[1]
		\Function{BackwardExt}{$(k,k',s),a$}
			\ForAll{$b<\overline{a}$}\Comment{$b$ can be $\$$}
				\State $k'\gets k'+\big[\pi(\overline{b},k+s)-\pi(\overline{b},k)\big]$
			\EndFor
			\State $s\gets \pi(a,k+s)-\pi(a,k)$
			\State $k\gets \pi(a,k)$
			\State \Return{$(k,k',s)$}
		\EndFunction
		\Function{ForwardExt}{$(k,k',s),a$}
			\State $(k',k,s)\gets${\sc BackwardExt}$((k',k,s),\overline{a})$
			\State \Return{$(k,k',s)$}
		\EndFunction
	\end{algorithmic}
\end{algorithm}

The \emph{double-strand suffix array interval} (\emph{ds-interval}) of $P$
is a 3-tuple $({\rm lo}(P),{\rm lo}(\overline{P}),{\rm occ}(P))$ with
${\rm lo}(P)\triangleq|\{i:T[i,n)<P\}|$ and ${\rm occ}(P)\triangleq|\{i:T[i,i+|P|)=P\}|$.
If the ds-interval of $P$ is $(k,k',s)$,
function {\sc BackwardExt()} in Algorithm~\ref{algo:backfor} calculates the ds-interval of $aP$
and {\sc ForwardExt()} calculates the ds-interval of $Pa$,
where $\pi(a,k)\triangleq |\{i:B[i]<a\}|+|\{i<k:B[i]=a\}|$ is the LF-mapping.
Please see \citet{Li:2024ac} for the proof of this algorithm.

4-tuple $(P,i,j,(k,k',s))$ is a $(\ell,c)$-match
if $(k,k',s)$ is the ds-interval of $P[i,j)$, $j-i\ge\ell$ and $s\ge c$;
it is a $(\ell,c)$-SMEM if there does not exist another $(\ell,c)$-match $(P[i',j'), (\tilde{k},\tilde{k}',\tilde{s}))$ such that $i'\le i<j\le j'$.
Intuitively, a $(\ell,c)$-match involves a $P$'s substring that is $\ge\ell$ in length and occurs $\ge c$ times in $T$;
a $(\ell,c)$-SMEM is a $(\ell,c)$-match that is not contained in other $(\ell,c)$-matches on the query string.
Section~\ref{sec:smem} will describe the algorithm to find $(\ell,c)$-SMEMs.

\subsection{The memory layout of BWT}

Minibwa and BWA-MEM encode the BWT the same way.
They split $B$ into blocks.
In each block, the first four 64-bit integers store the counts of {\tt A}/{\tt C}/{\tt G}/{\tt T} ahead of the block;
the next four 64-bit integers encode 128bp sequences in $B$ in the 2-bit encoding.
Each block takes 64 bytes both on disk and in memory.
When calculating $\pi(a,k)$, minibwa finds the address of the block that contains $B[k]$,
reads the counts at the beginning of the block and scans the encoded sequences to get the exact count up to $k$.
The BWT of the human genome takes $\sim$3 GB in RAM and is too large to fit the CPU cache.
Most calls to $\pi(a,k)$ incur cache misses.

\subsection{A brief introduction to prefetching}

\begin{figure}[b]
\includegraphics[width=\columnwidth]{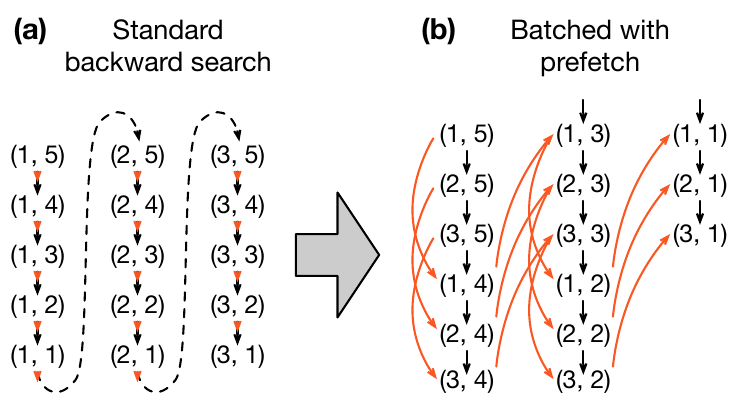}
\caption{Example of memory prefetch.
{\bf (a)} Standard backward search for three 5bp query strings.
$(r,i)$ denotes query $r$ at its position $i$.
Black arrows follow the computation order on a CPU.
Red arrows indicate data access in RAM.
{\bf (b)} Batched backward search with prefetch.}\label{fig:prefetch}
\end{figure}

A CPU takes hundreds of CPU cycles to access data in RAM but only several cycles to read data in the L1 cache.
If the data needed for computing is in RAM, the CPU has to wait for hundreds of CPU cycles before doing actual work.
Take the standard backward search as a case study.
With the classical algorithm, we process each query string in serial (Fig.~\ref{fig:prefetch}a).
Because the backward search at query position $(2,2)$ depends on the result on $(2,3)$,
the CPU has to wait for hundreds of CPU cycles to fetch data,
even though the computation itself only takes tens of CPU cycles approximately.

A more efficient approach is to decouple data access and computation by batching query strings (Fig.~\ref{fig:prefetch}b).
For example, after backward extension at $(2,3)$, we will know the address of data needed for computing $(2,2)$.
We can prefetch the data at this address, indicated by the red arrow from $(2,3)$ to $(2,2)$, and move onto position $(3,3)$.
When we come to $(2,2)$ following the black arrows, the data will be ready in cache and the CPU can perform backward extension of $(2,2)$ without waiting for the data.
This way we keep the CPU busy and effectively hide the latency of accessing data in RAM.

\subsection{Batched locate operation}

Suppose the ds-interval of $P$ is $(k,k',s)$.
$[k,k+s)$ only gives positions in the SA coordinate, while
$\{S(i):k\le i<k+s\}$ is the set of $P$'s positions in $T$.
To reduce memory with FM-index, we only store $S(i)$ when $i$ is a multiple of parameter $r$.
The ``locate'' operation finds the chromosomal positions of $[k,k+s)$ given a sampled suffix array.

Function {\sc LocateSlow()} in Algorithm~\ref{algo:sa} shows the canonical implementation of the locate operation which processes each SA position in turn.
Line 8 triggers data access and computation at the same time.
Function {\sc BatchLocate()} leverages the latency hiding technique.
Line 20 corresponds to black arrows in Fig.~\ref{fig:prefetch}b where the CPU does work;
line 21 corresponds to red arrows where prefetch happens.
Furthermore, when the input set $E$ consists of contiguous integers from a ds-interval,
line 20 will hit cache often and also greatly reduce cache misses.

To evaluate the performance of the two algorithms in Algorithm~\ref{algo:sa},
we filled $E$ with 20 random integers in range $[0,n)$.
On the FM-index of GRCh38, {\sc BatchLocate()} was over four times as fast as {\sc LocateSlow()} with identical output.
Latency hiding is an effective technique.

\begin{algorithm}[bt]
	\caption{Retrieve chromosomal positions}\label{algo:sa}
	\begin{algorithmic}[1]
		\LComment{$S$: suffix array with values only at a multiple of $r$}
		\LComment{$E$: set of suffix array positions}
		\Function{LocateSlow}{$S,r,E$}
			\State $F\gets\emptyset$
			\For{$k\in E$}
				\State $i\gets0$
				\While{$k\!\!\mod r\not=0$}
					\State $k\gets\pi(B[k],k)$
					\State $i\gets i+1$
				\EndWhile
				\State $F\gets F\cup\{i+S(k)\}$
			\EndFor
			\State \Return $F$\Comment{Set of chromosomal positions}
		\EndFunction
		\Function{BatchLocate}{$S,r,E$}
			\State $i\gets0; F\gets\emptyset$
			\While{$E\not=\emptyset$}
				\State $E'\gets\emptyset$
				\For{$k\in E$}
					\If{$k\!\!\mod r=0$}\Comment{$k$ is a multiple of $r$}
						\State $F\gets F\cup\{i+S(k)\}$
					\Else
						\State $l\gets\pi(B[k],k)$
						\State $E'\gets E'\cup\{l\}$
						\State Prefetch the memory block containing $B[l]$
					\EndIf
				\EndFor
				\State $i\gets i+1; E\gets E'$
			\EndWhile
			\State \Return $F$\Comment{Set of chromosomal positions}
		\EndFunction
	\end{algorithmic}
\end{algorithm}

\subsection{Finding SMEMs}\label{sec:smem}

Minibwa uses $(\ell,c)$-SMEMs to seed alignment.
Generalizing \citet{DBLP:conf/dlt/Gagie24} to the case of $c>1$,
\citet{Li:2024ac} provides an algorithm to find $(\ell,c)$-SMEMs given one query string.
This algorithm involves multiple rounds of forward and backward extensions (Algorithm~\ref{algo:backfor}).
For almost every extension, the CPU has to wait for the result of previous extension and then fetch the needed memory to cache.
The memory latency is the bottleneck.

Algorithm~\ref{algo:smem} alleviates this issue by batching query strings.
Overall, it uses a queue $Q$ to organize operations on all queries in a pipeline.
Same as the original serial algorithm~\citep{Li:2024ac},
the batched version tests if $P[i,i+\ell)$ is a match with backward extension (line 7--11 in stage 1).
If it is not a match, we jump and try again (line 32);
otherwise, we use forward extension to find the end of the SMEM starting at $i$ (line 12--21 in stage 2).
A second round of backward extension is applied to find the start of the next SMEM (line 22--26 in stage 3).
Minibwa prefetches memory blocks containing $B[l]$ and $B[l']$ after line 18 and 30.
It additionally precomputes the ds-intervals of all 10-mers to reduce the number calls to {\sc BackStep}().

\begin{algorithm}[tb]
	\caption{Find $(\ell,c)$-SMEMs in string set $\mathcal{P}$}\label{algo:smem}
	\begin{algorithmic}[1]
		\Function{BatchFindSMEM}{$\mathcal{P},\ell,c$}
			\ForAll{$P\in\mathcal{P}$}\Comment{initialize queue $Q$}
				\If{$\ell-1<|P|$}
					\State $Q.{\rm push}(1,P,0,\ell-1,(0,0,|B|))$
				\EndIf
			\EndFor
			\While{$Q$ is not empty}
				\State $(u,P,i,j,(k,k',s))\gets Q.{\rm shift}()$\Comment{take out the head}
				\If{$u=1$}\Comment{stage 1: backward extension}
					\If{$j<i$}\Comment{move to stage 2}
						\State $Q.{\rm push}(2,P,i,i+\ell,(k,k',s))$
					\Else
						\State {\sc BackStep}$(Q,u,P,i,j,(k,k',s),\ell,c)$
					\EndIf
				\ElsIf{$u=2$}\Comment{stage 2: forward extension}
					\If{$j=|P|$}\Comment{reach the end of $P$}
						\State Output $(P,i,j,(k,k',s))$\Comment{report an SMEM}
						\State {\bf continue}
					\EndIf
					\State $(l,l',t)\gets${\sc ForwardExt}$((k,k',s),P[j])$
					\If{$t\ge c$}\Comment{1-step forward ext.}
						\State $Q.{\rm push}(2,P,i,j+1,(l,l',t))$
					\Else\Comment{move to stage 3}
						\State Output $(P,i,j,(k,k',s))$\Comment{report an SMEM}
						\State $Q.{\rm push}(3,P,i,j,(0,0,|B|))$
					\EndIf
				\ElsIf{$u=3$}\Comment{stage 3: backward ext. again}
					\If{$j\le i$}\Comment{back to stage 1}
						\State $Q.{\rm push}(1,P,j+1,j+\ell,(0,0,|B|))$
					\Else
						\State {\sc BackStep}$(Q,u,P,i,j,(k,k',s),\ell,c)$
					\EndIf
				\EndIf
			\EndWhile
		\EndFunction
		\Function{BackStep}{$Q,u,P,i,j,(k,k',s),\ell,c$}\Comment{helper func.}
			\State $(l,l',t)\gets${\sc BackwardExt}$((k,k',s),P[j])$
			\If{$t\ge c$}
				\State $Q.{\rm push}(u,P,i,j-1,(l,l',t))$
			\ElsIf{$j+\ell<|P|$}
				\State $Q.{\rm push}(1,P,j+1,j+\ell,(0,0,|B|))$
			\EndIf
		\EndFunction
	\end{algorithmic}
\end{algorithm}

BWA-MEM uses an older algorithm to find SMEMs~\citep{Li:2012fk}.
On real short reads without batching, the minibwa algorithm is 10\% faster to find $(19,1)$-SMEMs due to the 10-mer cache.
With batching, minibwa is $\sim$2.5 times as fast.
Minibwa is $\sim$4.6 times as fast as BWA-MEM for finding $(19,2)$-SMEMs.
The gap is larger because Algorithm~\ref{algo:smem} skips short SMEMs more efficiently.

\subsection{Seeding, chaining and base alignment}\label{sec:align}

Minibwa seeds a query sequence $P$ in two rounds.
It first finds all $(19,1)$-SMEMs.
Suppose $(P,i,j,(k,k',s))$ is a resulting SMEM.
If $j-i\ge38$ and $s\le10$,
minibwa looks for $(\lfloor(j-1)/2\rfloor,s+1)$-SMEMs within $P[i,j)$.
During early development of BWA-MEM, we tried the same seeding algorithm.
However, as the old SMEM-finding algorithm is slow for $(19,2)$-SMEMs in the second round,
we ended up with a different seeding strategy in BWA-MEM.

Minibwa adapted the minimap2 chaining algorithm~\citep{Li:2018ab} for variable-length seeds.
It retains up to 50 best chains for a short read
and uses the minimap2 algorithm to close gaps between seeds or extends the first or the last seeds.
Minibwa first tries ungapped alignment.
If that alignment contains many mismatches, minibwa falls back to SIMD-based dynamic programming (DP) with dual-gap penalty~\citep{Suzuki:2018aa}.
For WGS short reads, the DP step is the performance bottleneck, taking $\sim$35\% of total CPU time.
Minibwa uses SSE4.1 intrinsics on x86 CPUs or NEON intrinsics on ARM CPUs.
We experimented an AVX2-based implementation but did not see clear improvement.

In comparison, BWA-MEM uses a heuristic chaining algorithm that is less accurate and does not work well for long reads.
It performs extension alignment from each seed but does not attempt to close gaps between seeds.
BWA-MEM uses the standard affine gap penalty and usually fills a smaller portion of the DP matrix given identical input sequences.
It is less tolerant with long gaps.
On the other hand, BWA-MEM tends to test more chains and still pays more effort overall for reads in centromeres or acrocentric short arms.

\subsection{Paired-end mapping}\label{sec:pe}

Minibwa finds properly paired reads using the same logic as BWA-MEM.
If a read is mapped but its mate is not mapped nearby,
minibwa locally rescues the mate in a window close to the mapped read.
On real data, this involves Smith-Waterman alignment~\citep{Smith:1981aa,Farrar:2007hs} between a $\sim$150 bp read $P$ and a reference substring $R$ of several hundred basepairs.
Unlike BWA-MEM that always initiates full alignment,
minibwa prefilters the read-reference pair by testing $q$-mer matches ($q=7$).
More exactly,
$$
M_t=|\{(i,j):\mbox{$P[i,i+q)=R[j,j+q)$ and $j-i=t$}\}|
$$
is the number of $q$-mer matches between $P$ and $R[t,t+|P|)$.
$\max_t M_t$ approximates the best ungapped alignment between $P$ and $R$.
Minibwa only initiates Smith-Waterman if $\max_t M_t\ge10$,
effectively reducing unsuccessful alignment.
It also applies mate rescue to fewer candidates in comparison to BWA-MEM.
Overall, minibwa might gain considerable performance at this step.

\subsection{Parameter setting based on read lengths}

BWA-MEM is primarily developed for short reads.
While minimap2 supports both short and long reads,
users have to specify a preset specific for either short or long reads but not both.
This is not ideal for mixed short and long reads.
Minibwa adjusts several parameters based on the read length $\ell$ using the following formula:
$$
\theta(\ell)=\theta_l-(\theta_l-\theta_s)\cdot 2^{-\frac{\max(\ell-\ell_{\rm min},0)}{\ell_{\rm mid}-\ell_{\rm min}}}
$$
where $\theta_l$ is the long-read value and $\theta_s$ is the short-read value.
We hardcode $\ell_{\rm min}=100$ and $\ell_{\rm mid}=2000$.
The length-adjusted parameter $\theta(\ell)$ reaches $(\theta_s+\theta_l)/2$ at $\ell=\ell_{\rm mid}$.

\subsection{Mapping directional bisulfite sequencing reads}\label{sec:meth}

The minibwa BS-seq read mapping algorithm is heavily influenced by BWA-Meth~\citep{Pedersen:2014aa} and BISCUIT~\citep{Zhou:2024ac}.
Like BWA-Meth, minibwa concatenates the C-to-T converted reference genome
and the G-to-A converted reference genome and constructs one FM-index for both strands.
This index consists of four copies of the reference genome:
forward C-to-T converted genome, forward G-to-A genome,
reverse G-to-A converted genome and reverse C-to-T genome.

Given a BS-seq read pair, minibwa converts all C bases on read 1 to T
and converts all G bases on read 2 to A.
It finds seeds against the converted FM-index.
A seed on read 1 is only retained if it comes from the forward C-to-T genome or the reverse G-to-A genome;
similarly, a seed on read 2 is retained if it comes from the forward G-to-A or the reverse C-to-T genome.
Due to base conversion, a reference T is considered a match to a read C, but this should not happen without SNPs.
To avoid such T-to-C mismatches, minibwa retrieves the original reference and the original read sequences in each seed match,
splits the seed match at T-to-C mismatches, and discards resulting split seeds shorter than 19 bp.
Minibwa applies chaining normally.
Like BISCUIT, it performs base alignment and mate rescue using the original reference and read sequences but with
an asymmetric scoring matrix that permits C-to-T mismatches and penalizes T-to-C mismatches.

\section{Results}

We evaluated the mapping accuracy on simulated reads and
measured the performance on real data.
We tested BWA-MEM v0.7.19~\citep{Li:2013aa}, BWA-MEM2 v2.2.1~\citep{DBLP:conf/ipps/VasimuddinMLA19},
BWA-MEM3 v0.2.1 (\url{https://github.com/fg-labs/bwa-mem3}),
Bowtie2 v2.5.5~\citep{Langmead:2012fk},
minimap2 v2.30~\citep{Li:2018ab},
strobealign v0.17.0~\citep{Sahlin:2022aa}
and minibwa-rs (\url{https://github.com/henriksson-lab/minibwa-rs}) for short reads,
rammap v1.0~\citep{Wang2026.05.26.726289} and
Winnowmap2 v2.03~\citep{Jain:2020aa,Jain:2022ua} for long reads
and BISCUIT v1.8.1~\citep{Zhou:2024ac}, Bismark v0.25.1~\citep{Krueger:2011aa}
and BWA-Meth v0.2.9~\citep{Pedersen:2014aa} for BS-seq reads.
Minibwa-rs is a Rust rewrite of minibwa that outputs identical alignment on several test datasets.
Most code was written by directly coding agents.

All tools were run on a Linux server equipped with two Xeon Gold 6130 CPUs at 2.10GHz.
We used the recommended setting of each tool except for applying option ``{\tt -b1 -i -I300,30}'' to BISCUIT
which would assume non-directional input and ignore read pairing by default.

\subsection{Mapping accuracy on simulated data}\label{sec:eval-sim}

\begin{figure*}
\includegraphics[width=\textwidth]{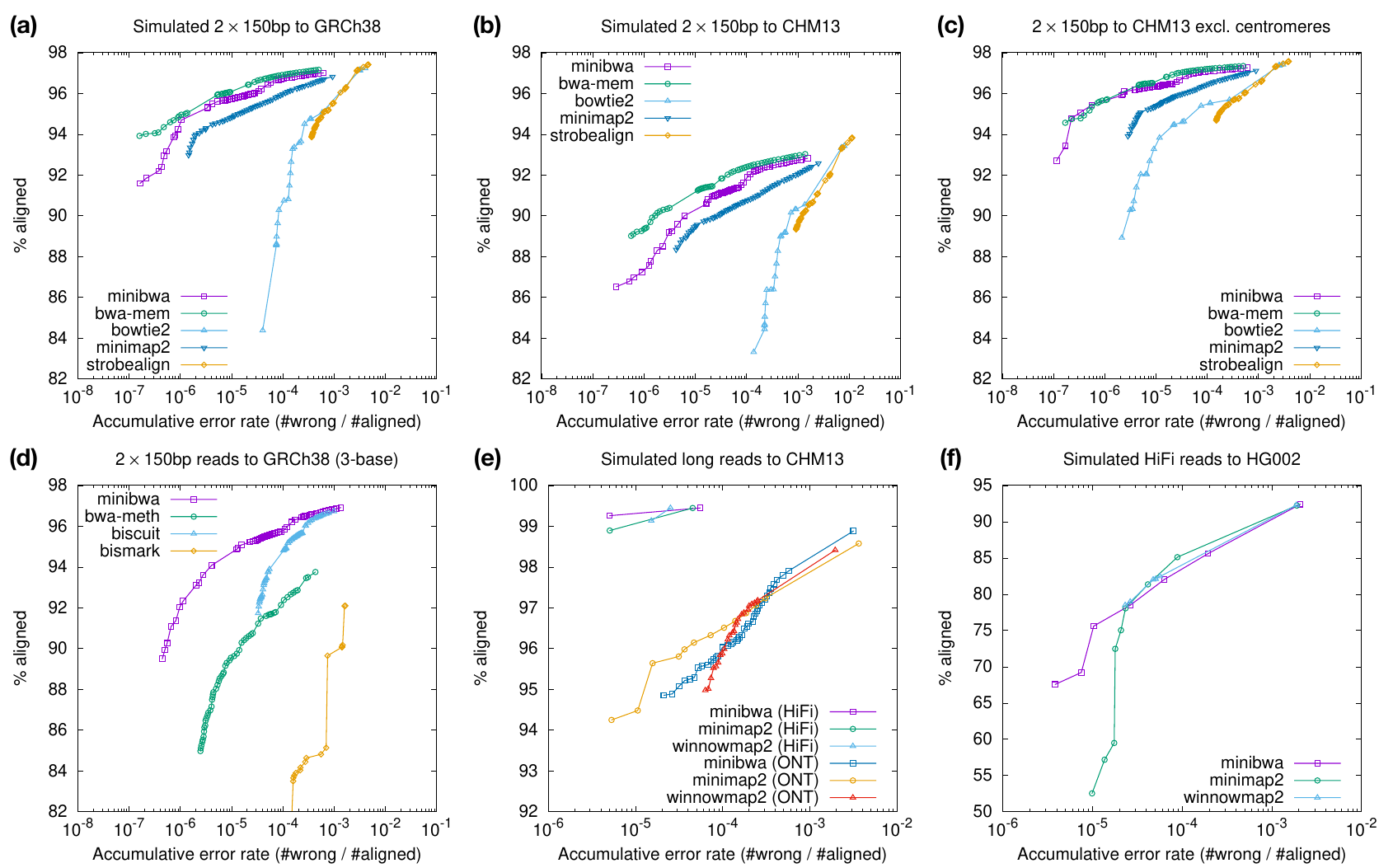}
\caption{Mapping accuracy measured on simulated reads.
Short and long reads are simulated from GRCh38, T2T-CHM13 or the HG002 diploid genomes and are mapped to the corresponding source genomes with various read aligners.
Suppose a read is originated from genomic interval $G_s$ in simulation
and is mapped to interval $G_a$ by an aligner.
It is considered mapped correctly if $|G_s\cap G_a|/|G_s\cup G_a|\ge10\%$.
Each point in the figures gives the fraction of mapped reads and the fraction of correctly mapped at a mapping quality threshold.
{\bf (a)} $2\times150$ bp paired-end reads simulated from GRCh38.
{\bf (b)} $2\times150$ bp paired-end reads simulated from T2T-GRCh38.
{\bf (c)} Same reads in {\bf (b)} but excluding reads from centromeres or acrocentric short arms.
{\bf (d)} Same reads in {\bf (a)} but mapped in the methylation mode.
{\bf (e)} HiFi and Nanopore reads simulated from T2T-CHM13.
{\bf (f)} HiFi reads simulated from the diploid HG002 genome consisting of two haplotypes.
}\label{fig:roc}
\end{figure*}

We simulated short reads using mason v2.0.9~\citep{fu_mi_publications962} with
``{\tt -{}-illumina-read-length 150 -{}-illumina-prob-mismatch-scale\\2.5}''.
We simulated long reads using Badread v0.4.2~\citep{Wick2019}
with the default option for Nanopore reads
and option ``{\tt -{}-error\_model pacbio2021 -{}-qscore\_model pacbio2021 -{}-length 15000,3000 -{}-identity 27,3}'' for HiFi reads.
We measured mapping errors by comparing the simulated position to the mapping position of each read (Fig.~\ref{fig:roc}),
stratified by mapping quality.
A mapper of higher accuracy tends to yield a curve closer to the upper left corner.

BWA-MEM is slightly more accurate than minibwa for WGS short-read alignment (Fig.~\ref{fig:roc}a and~\ref{fig:roc}b).
The difference is primarily driven by centromeric or acrocentric regions (Fig.~\ref{fig:roc}b versus~\ref{fig:roc}c)
as minibwa extends fewer reads and rescues fewer candidates in such regions (Section~\ref{sec:align} and~\ref{sec:pe}).
We will show later that the difference in theoretical accuracy does not translate into a difference in variant calling accuracy (Seciton~\ref{sec:eval-var}; Table~\ref{tab:var}).
Minibwa competes with minimap2 and Winnowmap2 for long-read mapping (Fig.~\ref{fig:roc}e and~\ref{fig:roc}f).

All BS-seq aligners in Fig.~\ref{fig:roc}d use the 3-base strategy and are therefore also capable of mapping normal WGS reads.
Both BWA-Meth and BISCUIT are built on top of BWA-MEM.
BWA-Meth is not as accurate due to its limitations mentioned in Introduction.
Bismark is known to map fewer reads than BWA-Meth~\citep{Gong_2022}.
It is unclear why BISCUIT is less accurate than minibwa despite using a similar strategy (Section~\ref{sec:meth}).
Notably, without ``{\tt -i -I300,30}'', BISCUIT would force reads to chromosomal contigs in GRCh38 and disable read pairing.
Its accuracy would be a lot lower, comparable to BWA-Meth.

We note that on real data, most false read alignment at high mapping quality is caused by large structural variants between individuals.
For example, if the reference genome has one copy of a gene but the sequenced sample has an extra copy,
all reads from the extra copy will be mismapped at high mapping quality.
The theoretical accuracy as is shown in Fig.~\ref{fig:roc} may not reflect accuracy in downstream data processing.
Nevertheless, such theoretical analysis helps to reveal subtle design flaws and hidden bugs in the implementation.
It is invaluable to the development of minibwa.

\subsection{Mapping performance on real data}

\begin{figure*}
\includegraphics[width=\textwidth]{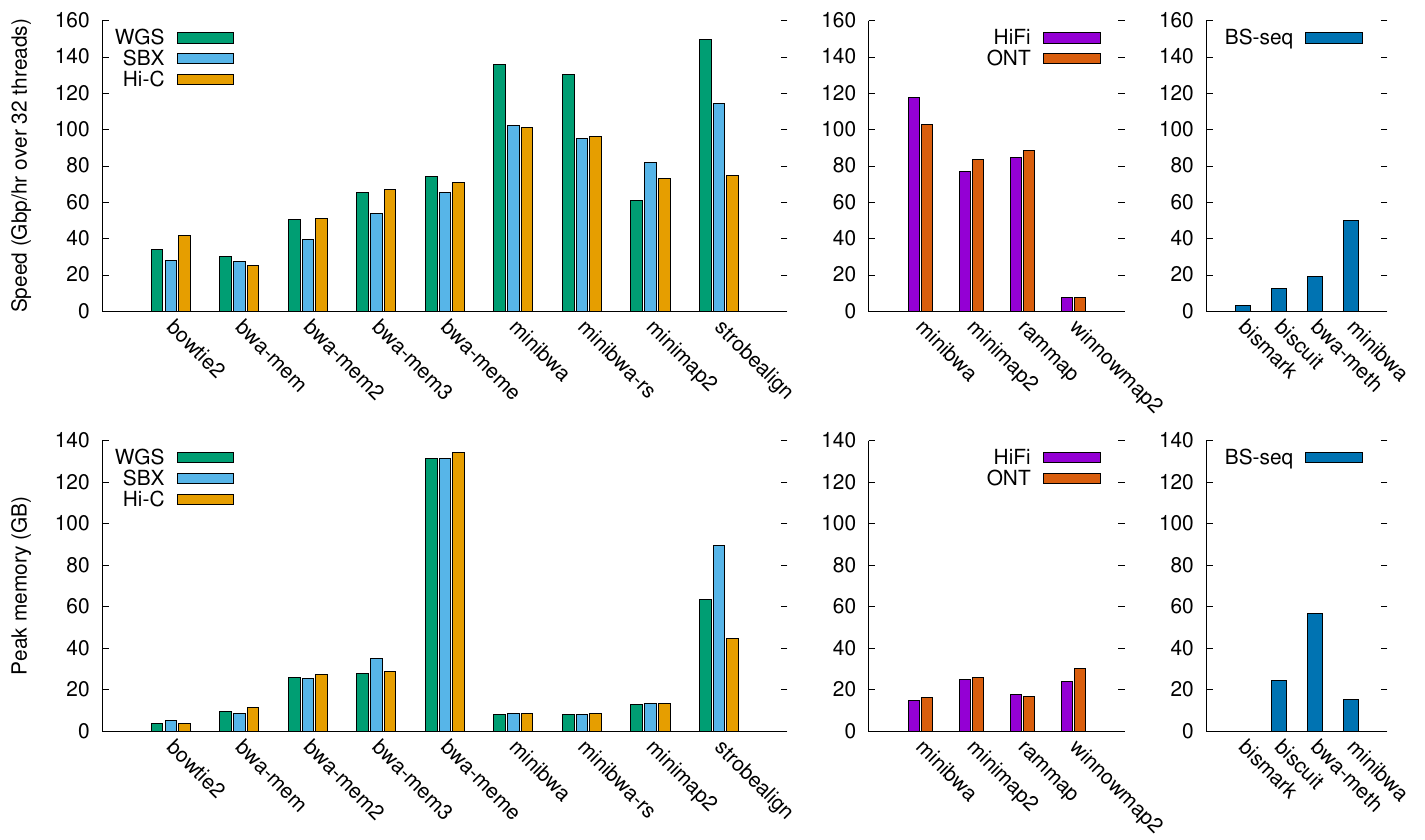}
\caption{Speed and peak memory on real data.
Reads are aligned to GRCh38.
32 CPU threads are specified for all mappers except for Winnowmap2 which automatically uses all available CPUs.
Alignment output is redirected to {\tt /dev/null} except for Bismark which has to generate BAM files.
The exact memory usage of Bismark could not be captured.}\label{fig:perf}
\end{figure*}

We obtained multiple whole-genome real datasets (see Data Availability) and aligned them to GRCh38 (Fig.~\ref{fig:perf}).
For standard WGS short reads, strobealign is the fastest.
BWA-MEM2 is 67\% faster than BWA-MEM and both BWA-MEME and BWA-MEM3 are even faster (Fig.~\ref{fig:perf}a).
Minibwa is about four times as fast as BWA-MEM and more than twice as fast as BWA-MEM2.
The speed advantage of minibwa over BWA-MEM2 narrows on Hi-C and SBX data where mate rescue is disabled.
Here, the N50 read length of this duplex SBX dataset is 243 bp, not much longer than the rest of short reads.
For long reads, minibwa is a little faster than minimap2 and both are an order of magnitude faster than Winnowmap2.
For BS-seq data, minibwa is a few times faster than BWA-Meth and BISCUIT and over 10 times faster than Bismark.
Across all six datasets, minibwa uses $<$20 GB memory at the peak, lower than most aligners except Bowtie2.

Minibwa-rs, the Rust rewrite, closely matches minibwa in speed for short reads
but for long reads, its relative performance to minibwa varies by $\sim$20\% across testing environments.
We thus did not show its timing in the figure.

\begin{table}[b]
\caption{Short-read variant calling accuracy with DeepVariant}\label{tab:var}
\tabcolsep=0pt
\begin{tabular*}{\columnwidth}{@{\extracolsep{\fill}}lrrrr@{\extracolsep{\fill}}}
\toprule
& \multicolumn{2}{@{}c@{}}{SNP} & \multicolumn{2}{@{}c@{}}{Indel} \\
\cline{2-3}\cline{4-5}
Aligner & \#FN & \#FP & \#FN & \#FP \\
\midrule
minibwa     & {\bf 46,367} & {\bf 7,544} & {\bf 36,321} & 5,308 \\
BWA-MEM     & 46,895 & 7,585 & 37,425 & {\bf 5,218} \\
strobealign & 57,545 & 8,108 & 37,828 & 5,655 \\
\botrule
\end{tabular*}
\begin{tablenotes}\setlength\itemsep{0.0em}
FN: false negative; FP: false positive.
Numbers in bold indicate the best performer on each metric.
\end{tablenotes}
\end{table}

\subsection{Effect on short-read variant calling}\label{sec:eval-var}

We aligned HG002 short reads~\citep{Baid2020.12.11.422022} to GRCh38, called small variants with DeepVariant v1.10.0~\citep{Poplin:2018ab}
and compared the calls to the GIAB Q100 ground truth v1.1~\citep{Hansen2025.09.21.677443} with RTG vcfeval v3.13~\citep{Cleary023754}.
Minibwa closely matched or slightly improved on BWA-MEM (Table~\ref{tab:var}):
528 fewer SNP FNs, 41 fewer SNP FPs, 1,104 fewer indel FNs, and 90 more indel FPs.
We inspected BWA-MEM-specific indel FNs and found minibwa tends to align more reads with $>$10 bp indels,
consistent with the algorithmic analysis (Section~\ref{sec:align}).
Although strobealign is 10\% faster than minibwa (Fig.~\ref{fig:perf}), it does not compete with minibwa or BWA-MEM on small variant calling.

\section{Discussions}

We developed minibwa for aligning genomic short and long reads and bisulfite sequencing reads.
It trades full compatibility with BWA-MEM for deeper algorithmic improvements.
Although minibwa is not bit-identical to BWA-MEM in its output,
it produces equivalent or slightly improved small variant calls,
and more importantly, enables long-read alignment and doubles performance over BWA-MEM2.
As the developers of BWA-MEM and minimap2, we consider minibwa the full replacement of BWA-MEM.
The only BWA-MEM feature missing from minibwa is the support of alternate contigs in the reference genome
which is not well implemented in BWA-MEM.
We plan to redesign this feature for improved accuracy in future releases.

\section*{Acknowledgments}

We are grateful to Johan Henriksson for porting minibwa to Rust at \url{https://github.com/henriksson-lab/minibwa-rs}.

\section*{Author contributions}

H.L. conceived the project, implemented the core algorithm, analyzed the data and drafted the manuscript.
N.H. added new features, contributed to development and reviewed the draft.

\section*{Conflict of interest}

N.H. is an employee of Fulcrum Genomics LLC.

\section*{Funding}

This work is supported by National Institute of Health grant R01HG010040, R01HG014175 and U24CA294203 (to H.L.).

\section*{Data availability}

The minibwa source code: \url{https://github.com/lh3/minibwa};
HG002 NovaSeq short reads: \url{https://bit.ly/hg002-baid-sr-30x};
HG002 PacBio HiFi reads: \url{https://bit.ly/hg002-hprc-revio};
HG002 Nanopore R10 reads: \url{s3://ont-open-data/giab\_2023.05/};
HG002 Hi-C short reads: \url{https://bit.ly/hg002-hprc-hic1};
HG002 SBX reads: \url{https://bit.ly/hg002-sbx-091025};
NA12878 directional bisulfite sequencing reads: SRR4235788.

\bibliographystyle{apalike}
{\sffamily\small
\bibliography{minibwa}}

\end{document}